\documentclass[]{spie}

\usepackage{amsmath,amsfonts,amssymb}
\usepackage{graphicx}
\usepackage[colorlinks=true, allcolors=blue]{hyperref}

\title{Designing a Photonic Physically Unclonable Function Having Resilience to Machine Learning Attacks}

\author[a]{Elena R. Henderson}
\author[a]{Jessie M. Henderson}
\author[a]{Hiva Shahoei}
\author[b]{William V. Oxford}
\author[a]{Eric C. Larson}
\author[a]{Duncan L. MacFarlane}
\author[a]{Mitchell A. Thornton}
\affil[a]{Darwin Deason Institute for Cyber Security, Southern Methodist University, \hspace{7em} 6425 Boaz Lane Dallas, TX 75205, USA}
\affil[b]{Anametric, Inc., Austin, TX 78652, USA}

\authorinfo{Send correspondence to ERH: erhenderson@smu.edu. Further author information: JMH: hendersonj@smu.edu, HS: hshahoei@smu.edu, WVO: oxford@anametric.com, ECL: eclarson@smu.edu, DLM: dmacfarlane@smu.edu, MAT: mitch@smu.edu}

\begin{document} 
\maketitle

\begin{abstract}
Physically unclonable functions (PUFs) are designed to act as device `fingerprints.'
Given an input challenge, the PUF circuit should produce an unpredictable response for use in situations such as root-of-trust applications and other hardware-level cybersecurity applications.
PUFs are typically subcircuits present within integrated circuits (ICs), and while conventional IC PUFs are well-understood, several implementations have proven vulnerable to malicious exploits, including those perpetrated by machine learning (ML)-based attacks.
Such attacks can be difficult to prevent because they are often designed to work even when relatively few challenge-response pairs are known in advance.
Hence the need for both more resilient PUF designs and analysis of ML-attack susceptibility.
Previous work has developed a PUF for photonic integrated circuits (PICs).
A PIC PUF not only produces unpredictable responses given manufacturing-introduced tolerances, but is also less prone to electromagnetic radiation eavesdropping attacks than a purely electronic IC PUF.
In this work, we analyze the resilience of the proposed photonic PUF when subjected to ML-based attacks. 
Specifically, we describe a computational PUF model for producing the large datasets required for training ML attacks; we analyze the quality of the model; and we discuss the modeled PUF's susceptibility to ML-based attacks.
We find that the modeled PUF generates distributions that resemble uniform white noise, explaining the exhibited resilience to neural-network-based attacks designed to exploit latent relationships between challenges and responses.
Preliminary analysis suggests that the PUF exhibits similar resilience to generative adversarial networks, and continued development will show whether more-sophisticated ML approaches better compromise the PUF and---if so---how design modifications might improve resilience.
\end{abstract}


\section{Introduction}\label{sec:intro}
Physically unclonable functions, or PUFs, act as unique identifiers---or `fingerprints'---for devices \cite{pappu02}.
PUFs accept inputs termed \textit{challenges} and produce outputs termed \textit{responses}, and the relationship between a challenge and response is informed by probabilistic geometric anomalies that result from fabrication imperfections on integrated circuits\cite{pappu02,naveenkumar22,mahalat19,sahoo15,ruhrmair10}.
Ideally, these imperfections produce challenge-response pairs (CRPs) in which the response is not predictable given the challenge, and in which different PUF instances produce different responses for the same challenge.
Several conventional PUF architectures\cite{maes13,maiti11,mcgrath19,naveenkumar22,mahalat19,sahoo15,lim05,katzenbeisser12} have been applied to situations such as device authentication, trusted system development, and hardware-level cybersecurity\cite{guajardo07,lofstrom00,pappu02,gassend02,gassend02a,lee04}.

Machine learning (ML)-based attacks that predict responses for as-of-yet-unseen challenges have become an increasing threat to PUFs of all variants\cite{delvaux19,hazari21,kumar18,ge20,santikellur19,machida15,aghaie21,khalafalla19,saha13,laguduva20,guo16,ruhrmair10,van13,yoon20,mishra23,ruhrmair13,dodda21}.
These attacks do not directly represent the transformation between challenges and responses, but rather predict the result of that transformation after learning from a set of CRPs collected from a given PUF.
Machine-learning-based attacks, including regression mappings, state vector machines, $k$-nearest neighbors, random forests, genetic algorithms, artificial neural networks, and generative adversarial networks, have proven effective against various conventional PUFs, and additional work has developed theoretical relationships between the number of CRPs used for training and response prediction accuracy \cite{ruhrmair14,delvaux19,ge20,santikellur19,machida15,khalafalla19,aghaie21,hazari21,kumar18,saha13,ganji16,yoon20,laguduva20,guo16,siddhanti19,ganji16a,ganji15}.

Consequently, there is reason to explore alternative PUF structures, of which photonic PUFs are a promising candidate.
Photonic PUFs leverage sensitive manufacturing tolerances of photonic integrated circuit (PIC) components to produce CRPs that are difficult to predict, in part because accurately representing the function that maps challenges to responses may require infinitely many matrix-vector calculations \cite{macfarlane23,jones42}.
Because the relationship between the inputs and outputs is highly nonlinear, photonic PUFs have the potential to exhibit lesser susceptibility to ML-based threats than do some conventional PUFs\cite{pavanello21}.
Additionally, it is more difficult for an attacker to collect the CRPs required for training ML-based attacks: the use of light as input and output makes photonic PUFs less prone to physical and electromagnetic radiation eavesdropping attacks\cite{helfmeier14}.
Previous work developed a photonic PUF architecture\cite{macfarlane23} that we have shown exhibits resilience to some ML-based threats\cite{henderson24}.
The goal of this work, then, is to analyze that PUF's properties to better explain this demonstrated resilience.

The rest of the paper is as follows: Section \ref{sec:background} provides a brief introduction to PUF-quality metrics and to the class of optical/photonic PUFs.
Section \ref{sec:modeling_the_puf} then describes the photonic PUF and---specifically---a computational model thereof that allows for obtaining large synthetic datasets far more easily than from fabricated photonic integrated circuit PUFs.\footnote{While this work will not directly consider results from fabricated PIC PUFs, we note that preliminary experimental measurements from a fabricated PUF circuit indicate good correspondence with our model.}
Then, Section \ref{sec:results} analyzes ten synthetic datasets (\textit{i.e.}, ten PUFs) from the computational model, and illustrates the properties that make the PUF less susceptibile to ML-based threats \cite{henderson24}.
Finally, we close by contextualizing our findings and directions for further research.

\section{Background}\label{sec:background}

\subsection{Brief Introduction to Physically Unclonable Functions}
For over twenty years, physically unclonable functions have been studied for cybersecurity applications\cite{pappu02}.
This has led to several proposed architectures\cite{maes13,maiti11,mcgrath19,naveenkumar22,mahalat19,sahoo15,lim05,katzenbeisser12}, and while a review of these designs is beyond the scope of this work, it is worth introducing a classification based upon PUF behavior.
\textit{Strong} PUFs are defined as being impossible to clone, being impossible to characterize completely within a limited time frame, and having responses that are difficult to numerically predict for any given challenge\cite{ruhrmair10,ruhrmair14}.
By contrast, \textit{weak} PUFs have a limited number of challenges, and their responses are not designed for direct transmission outside of the device\cite{ruhrmair10,ruhrmair14}.
Because these definitions are qualitative, several properties have been developed to quantitatively evaluate PUF strength.
The three most oft-discussed PUF properties are uniqueness, uniformity, and reliability, and other similar characteristics include bit aliasing and susceptibility\cite{ruhrmair10,ruhrmair14,aseeri18}.
The remainder of this section introduces these properties, which are computed using sets of CRPs, which are usually represented as bitstrings.
So, this work's analysis assumes that challenges and responses are comprised of bits.

Uniqueness is an inter-PUF measure of the similarity between the responses produced by a challenge sent into several PUFs\cite{aseeri18}.
Uniqueness is a number between zero and one, inclusive; it is zero when the responses are exactly the same and one when they are inverted versions of each other.
The ideal uniqueness is $0.5$, which indicates that $50\%$ of the bits in the response corresponding to any given challenge vary between PUFs.
When measuring across $k$ fabricated PUFs, each with $n$-bit responses and $m$ CRPs, the uniqueness is computed as $\phi_{unique} = \frac{2}{k(k-1)m}\sum_{l=1}^m\sum_{i=1}^{k-1}\sum_{j=i+1}^k \frac{H(R_{il},R_{jl})}{n}$, where $H(\cdot,\cdot)$ is the Hamming distance between the arguments, and $R_{il}$ and $R_{jl}$ are the responses for chips $i$ and $j$ with CRP index $l$.

Uniformity is an intra-PUF measure of how many zeros and ones appear overall in bitstring responses for a given PUF\cite{aseeri18}.
Uniformity is also a number between zero ane one, inclusive; it is zero when every response bit is zero, and it is one when every response bit is one.
The ideal uniformity is $0.5$, indicating that half of the overall response bits are zero and the other half are one.
When measuring for a given PUF across $m$ CRPs with $n$-bit responses, the uniformity is computed as $\phi_{uniformity} = \frac{1}{n*m}\sum_{l=1}^m\sum_{i=1}^n r_{il}$, where $r_{il}$ is the $i$th bit of the $l$th CRP.

Reliability is an intra-PUF measure of how often a given challenge produces the same response for a given PUF.
It also varies between zero and one, inclusive; a reliability of zero indicates that a given challenge always produces the same response when input into a certain PUF, while a reliability of one indicates that a repeated challenge's response is entirely inverted when sent into a PUF multiple times.
The ideal reliability is thus zero, because a PUF should provide an identifying `fingerprint' that consistently reproduces the same response when prompted with a given challenge.
When measuring for a PUF at $k$ different times (and thus with as many as $k$ different sets of operating conditions), $n$-bit responses, and $m$ CRPs, the reliability is computed as $\phi_{reliability} = 1 - \frac{1}{k*m}\sum_{l=1}^k\sum_{i=1}^m \frac{H(R_i,R_{il})}{n}$, where $H(\cdot,\cdot)$ is the Hamming distance between the arguments, $R_i$ is the $i$th response of a `baseline' set of $m$ responses, and each $R_{il}$ is the $i$th response in a set of $m$ responses for the same challenges, but for a different data collection scenario (\textit{i.e.}, different value of $l$).

Bit aliasing is both an intra- and inter-PUF measure, because it can be used both within and across PUFs.
It measures the distribution of zeros and ones for a given bit position of the responses, meaning it takes values between zero and one, inclusive.
A bit aliasing of zero indicates that the bit at a given position in the responses always takes a value of zero, while a bit aliasing value of one indicates that the bit at a given position in the responses always takes a value of one.
Thus, the ideal value is $0.5$, indicating that a bit position in the responses varies evenly between zero and one.
When measuring across $k$ PUFs with $n$-bit responses and $m$ CRPs, the bit aliasing is given by $\phi_{ba,bp} = \frac{1}{k*m}\sum_{l=1}^k\sum_{i=1}^m r_{il}$, where $r_{il}$ is the bit at index $bp$ ($1$ through $n$) for the $i$th response of the $l$th PUF.

Finally, susceptibility measures how likely it is that a PUF will be compromised.
Given the increasing number of available ML-based attacks, this has become an oft-discussed metric\cite{aseeri18,delvaux19,ge20,santikellur19,machida15,khalafalla19,aghaie21,hazari21,kumar18,siddhanti19,ganji16,yoon20}, but it may have different meanings in different contexts.
For example, it might include the `physical' susceptibility of a PUF, meaning how well CRPs can be detected using a side channel attack\cite{helfmeier14,mispan21}.
Or, it might represent how many CRPs are required to train an ML-based attack that can predict responses better than chance\cite{delvaux19,hazari21,kumar18,ge20,santikellur19,machida15,aghaie21,khalafalla19,aseeri18,siddhanti19,ganji16,laguduva20,guo16}.
Because a primary rationale for the photonic PUF is its lesser vulnerability to physical CRP-collection attacks, we assess susceptibility as the number of CRPs required to launch an ML-based attack that can predict responses for as-of-yet-unseen challenges with a probability greater than chance.
This susceptibility value is informed by properties such as uniqueness, uniformity, and bit aliasing.

Section \ref{sec:results} analyzes these properties for the modeled photonic PUF, but before turning to a description of that model, we briefly introduce the class of photonic---or optical---physical unclonable functions.

\subsection{Photonic Physically Unclonable Functions}
Because PUFs that use light-encoded data as challenge and response are less susceptible to side-channel threats that collect CRPs, previous work has studied several optical/photonic PUF architectures\cite{pavanello21,pavanello23,marakis22,lio23,nocentini22,akriotou20,tarik22,grubel17,grubel17a,grubel18,knechtel19,macfarlane23}.
Designs include optical scattering PUFs, which use the movement of scattered light around nano-sized obstacles to obtain responses\cite{marakis22,lio23,pavanello23,nocentini22}, and ``silicon'' photonic PUFs that use the interaction of light in a ``chaotic silicon microcavity'' to do the same\cite{grubel17,grubel17a,tarik22}.
Other designs control light's traversal through PIC components\cite{pavanello23,macfarlane23}, and Ref.~\citenum{macfarlane23} introduced a PUF of this type that is considered in this work.
Experimental characterization of that PUF's components suggests that they exhibit fabrication-imposed geometric idiosyncrasies for providing an unpredictable nonlinear relationship between challenges and responses\cite{macfarlane23,shahoei22}.
The goal of this paper is to further support that finding by analyzing properties of these photonic PUFs on a larger scale.
Previous characterization has been solely for individual PUF components with a few CRPs, so we seek to analyze hundreds of thousands of CRPs per PUF for several PUFs.
Collecting this much data from fabricated photonic PUFs is challenging for multiple reasons, including the substantial expense of PIC PUF fabrication.
Consequently, the next section describes a computational model of the photonic PUF that efficiently generates hundreds of thousands of CRPs for several PUFs.

\section{Modeling the Photonic PUF}\label{sec:modeling_the_puf}
\begin{figure} [ht]
   \begin{center}
   \begin{tabular}{c}
   \includegraphics[height=5cm]{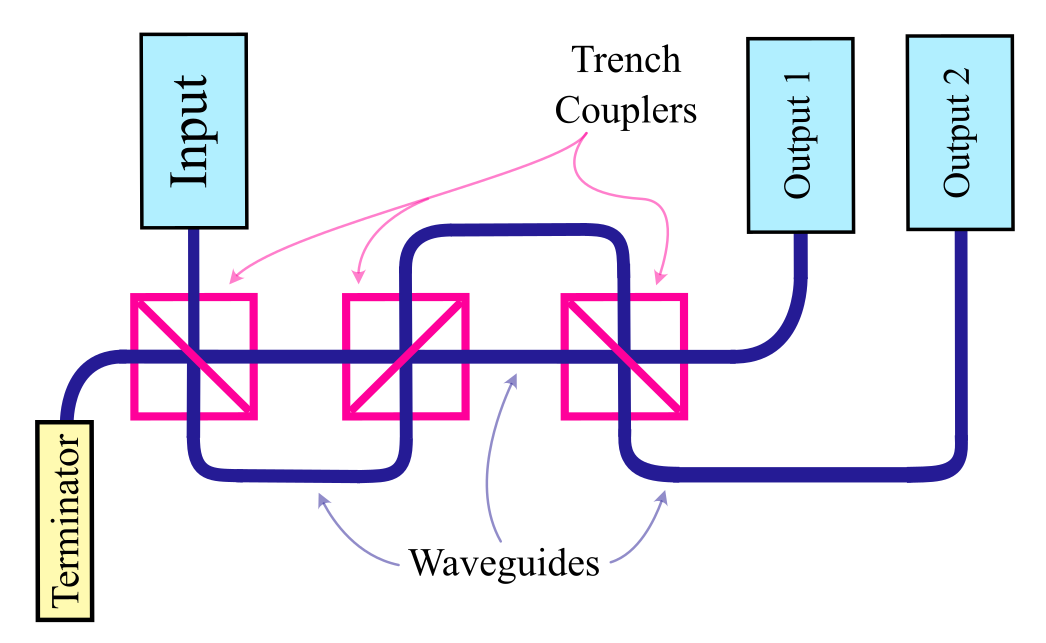}
   \end{tabular}
   \end{center}
   \caption
   { \label{fig:photonic_puf_cell} 
The photonic PUF cell. It is comprised of three trench couplers, connected by waveguides. There are two outputs, which are used to create the final outputs from interim cells, as shown in Figure \ref{fig:photonic_puf_architecture}.}
\end{figure}

There are four primary aspects of the photonic PUF model: the structure of a cell, the structure of the PUF architecture and some ramifications thereof, the mathematical representation of the components, and the mathematical representation of the challenges and responses.
We begin with the structure of a cell, which comes directly from Ref.~\citenum{macfarlane23}.
As illustrated in Figure \ref{fig:photonic_puf_cell}, an input is connected to a series of waveguides that connect three trench couplers, ultimately producing two outputs.
The inputs and outputs are beams of light, and properties of the light---when represented in bitstring form---comprise the challenges and responses.\footnote{In a laboratory setting with a fabricated device, there are several properties of the input and outputs that can be used to create CRPs; we will discuss how the model represents CRPs later in this section.}

\begin{figure} [ht]
   \begin{center}
   \begin{tabular}{c}
   \includegraphics[width=0.8\linewidth]{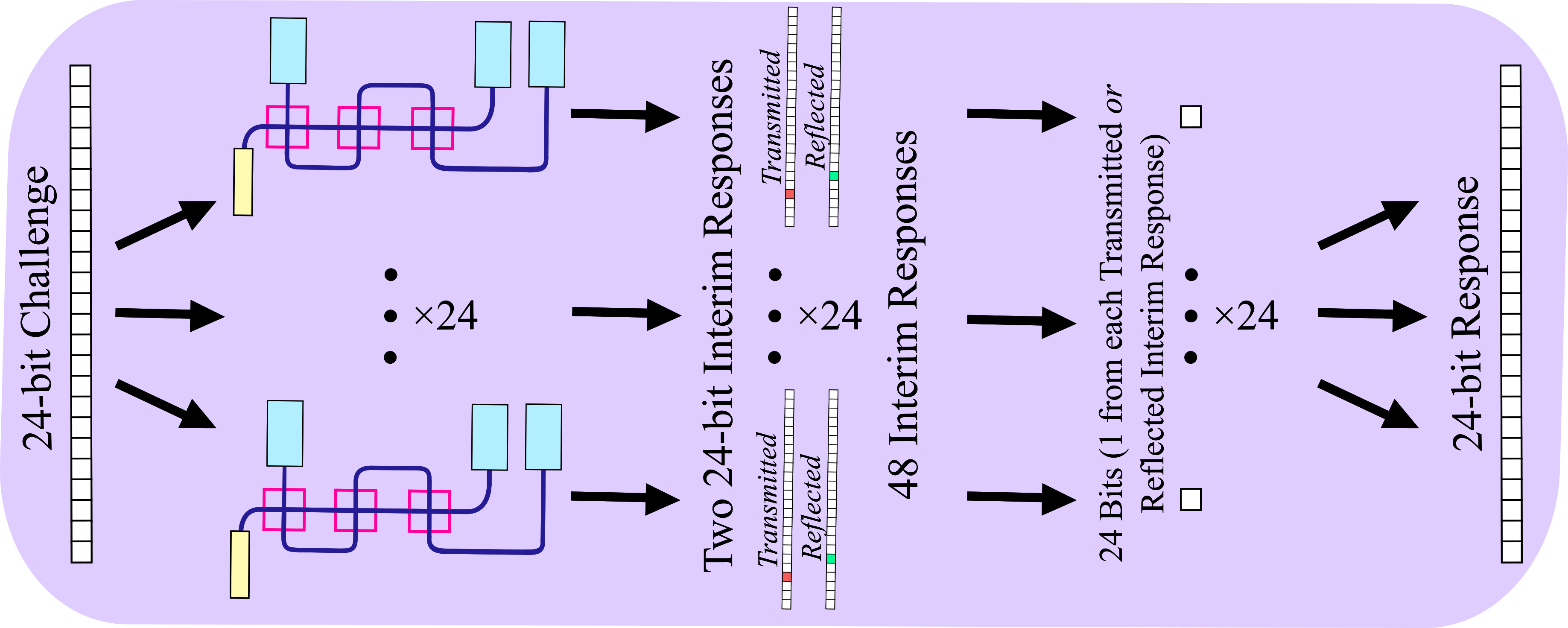}
   \end{tabular}
   \end{center}
   \caption
{ \label{fig:photonic_puf_architecture} 
The photonic PUF. A single challenge is sent to each of $24$ cells, which each produce two interim responses. Each of those interim responses has $24$ bits, and the final response for the given challenge is made from one bit of each of either Output $1$'s or Output $2$'s interim responses. For example, we might create a final response by combining the bits at index three (highlighted in red) of Output $1$'s interim responses, or by combining the bits at index five (highlighted in green) of Output $2$'s interim responses.}
\end{figure}

These cells are combined to produce the overall PUF architecture.
As illustrated in Figure \ref{fig:photonic_puf_architecture}, the overall responses are comprised of one bit from each interim cell response.
As will be demonstrated in Section \ref{sec:results}, this cell-based architecture lends responses that are not highly correlated, meaning the challenge-response distribution is more uniform and it should be harder to predict responses even with a set of known CRPs.
The combination of interim response bits to create final bits lends many possible final PUF responses, which we term \textit{PUF interpretations}.
For example, Figure \ref{fig:photonic_puf_architecture} illustrates that we might create a PUF interpretation by combining the fourth bit of Output $1$'s interim responses (highlighted in red), or we might create a PUF interpretation by combining the sixth bit of Output $2$'s interim responses (highlighted in green).
In Section \ref{sec:results}, we consider 48 interpretations per PUF to assess the effect of the interim bit used to create the overall responses on PUF quality.
Specifically, as illustrated in Figure \ref{fig:photonic_puf_architecture}, we use $24$-bit interim responses to create $24$-bit PUF interpretations, we do not mix interim bits from Output $1$ and Output $2$ when constructing PUF interpretations, and we construct PUF interpretations using one bit of the same index from each cell.
Therefore, we analyze $24$ PUF interpretations for Output $1$ and also for Output $2$, lending a total of $48$ interpretations per PUF.

Before turning to the mathematical representation of the circuit, we make three notes about this architecture.
First, there is no reason that the number of cells must necessarily equal the number of bits in the challenges and responses.
We choose this structure to analyze PUF interpretations based on the bit index used---and not the number of bits from each cell used.
Consequently, we consider the `limiting case' of interpretations made with just one bit from each cell.
Second, and again to limit the number of cases for analysis, we choose not to mix interim response bits from Outputs $1$ and $2$.
Preliminary experimental results from a fabricated PIC PUF suggest that it is possible to use bits from both Output $1$ and Output $2$ to create a single PUF interpretation, but exploring the effects of such mixing---especially on the response correlation---is left for future work.
Third and finally, if PUF interpretations were constructed using bits from different bit indices (instead of, say, the fifth bit from each interim response), there would be far more interpretations per PUF.
While we limit our scope to only the $48$ PUF interpretations described above, exploring mixed bit indices is another avenue of further research.

We now move to the core of the model: Jones calculus is a method for modeling the behavior of optical components using $2\times2$ Jones matrices to represent each component\cite{jones42,brosseau98}.
Lossless Jones matrices are unitary, have complex numbers as entries, have determinants less than or equal to one, and have traces less than or equal to two.
In our computational model, each waveguide and trench coupler is assigned a Jones matrix with randomly-selected parameters that are checked to ensure the above constraints are satisfied.
Each component's Jones matrix then represents the unique behavior of that component, including any fabrication imperfections.

Jones calculus also provides a representation for a state as it moves through the PUF.
The challenges and responses in non-bitstring form are represented as \textit{Jones vectors}, which represent the state of light as $\begin{bmatrix}
E_xe^{i\phi_x} \\ E_ye^{i\phi_y}
\end{bmatrix}$, where $E_x$ and $E_y$ represent the field magnitudes of the state, and $\phi_x$ and $\phi_y$ are the phases.
Our computational model allows for challenges specified as Jones vectors, and also allows for automated production of Jones vectors with evenly-spaced field magnitudes and phase differences ($\Delta\phi=\phi_y-\phi_x$).
Thus, the model can efficiently produce CRP datasets that include challenges spaced over an entire domain specified by ranges for $E_x$, $E_y$, $\phi_x$, and $\phi_y$.

For the datasets in this work, we take $E_x, E_y \in (0,1)$ and $\phi_x,\phi_y \in [0,2\pi)$, with step sizes of $\Delta E_x^2=0.0003$ and $\Delta \phi_y = 0.087$ radians to accommodate the precision limitations of laboratory devices.
Using a specified $E_x^2$ and $\phi_y$, the model can construct the other parts of a challenge vector: set $E_y = \sqrt{1-E_x^2}$ and let the phase difference $\Delta\phi=\phi_y\Rightarrow\phi_x=0$.
(This is an acceptable arrangement, because it will generate challenge vectors that span the entire $\Delta\phi$ range, and it is more computationally efficient than varying both $\phi_y$ and $\phi_x$ to create challenges.) 

We then represent the challenge bitstring as $24$ bits comprised of two $12$-bit parts: one part is the $E_x^2$ magnitude, and the other is the phase difference, $\Delta\phi$.
There are $12$ bits for each part, because that allows for representing the challenges without bitstring duplication.
Specifically, because we do not need a non-zero integer part for the $E_x^2$ value (we approach $E_x^2=1$, but we never reach it), we have $2^{-12} \approx 0.0002$, which can accommodate step sizes of $0.0003$.
Similarly, a three-bit integer part (to accommodate the $5$ and $6 < 2\pi$) leaves $2^{-9}\approx0.00195$, which can accommodate step sizes of $0.087$ radians.

Running the challenges in Jones vector form through our computational model results in a set of response Jones vectors, which can be converted to bitstrings using the same approach as for challenge bitstrings.
Unlike with the challenges, we cannot guarantee that all the response bitstrings will differ, because it is possible that the response Jones vectors will have $E_x^2$ and $\Delta\phi$ values that are smaller than $0.0003$ and $0.087$, respectively.
And because we cannot know \textit{a priori} what bit resolution will be required to uniquely represent all of the responses, we choose the same bit resolution as for the challenges.
Analysis has shown that this leads to PUF with robust properties (see Section \ref{sec:results}), so we leave experiments with different response bitstring lengths to future work.
After converting each of the interim responses to bitstrings, we create PUF interpretations by simply combining bits from the same output type (\textit{i.e.}, Output $1$ or $2$) and a given bit index (\textit{i.e.}, zero through $23$).
These stepsize and bitstring parameters provide a total of $212$,$929$ CRPs per PUF; the challenges are fixed across all PUFs, while the responses of course vary by PUF.

Using this PUF model, we collect data for ten different PUFs, and analyze the properties discussed in Section \ref{sec:background} with one exception.
We do not analyze reliability, because the computational model is deterministic for a given PUF; the model does not adjust the challenge-to-response-conversion based upon environmental considerations such as ambient temperature.
Consequently, it is not sensible to report on the reliability of synthetic datasets from the model.
We note that preliminary experimental measurements from a fabricated PUF circuit indicate good correspondence with our model.

\section{Analysis of Synthetic Photonic PUFs}\label{sec:results}
\subsection{Motivation and Data Collection}
Analysis of the photonic PUF began with susceptibility.
As described in Section \ref{sec:background}, susceptibility is a measure of how easily a given PUF may be compromised, and this paper defines susceptibility as the number of CRPs that are required to train an ML model to predict response bits better than chance.
We found that the photonic PUF demonstrates resilience to various neural-network based attacks, requiring $O(10^3)$ CRPs to train a model that can predict responses any better than chance, and requiring $O(10^4)$ CRPs to train a model that can predict responses with at least $65\%$ accuracy\cite{henderson24}.

After demonstrating the modeled photonic PUF's resilience to neural-network-based attacks, we began investigating the PUF's resilience to other forms of ML-based attack, including those from generative adversarial networks (GAN).
GANs are unsupervised models that learn not by benchmarking against pre-classified data, but rather by two networks competing against each other, with one network attempting to fool the other into classifying invented data as part of a given dataset\cite{goodfellow20}.
The two component models in a GAN are called the generator and the discriminator, and while the generator uses information from a given dataset to produce values that `fit' the dataset, the discriminator attempts to classify the generator's output as either a `true' member of the given dataset or as a `fake' invented output.
The generator's goal is to learn the statistical representation of the given dataset such that it can generate values that the discriminator erroneously classifies as members of the dataset.
Thus, a GAN attack against the photonic PUF attempts to replicate the relationship between challenges and responses by generating responses that fit the statistical `shape' of the response landscape.
When preliminary experiments found that a straightforward GAN was no better at predicting responses than were the neural network-based attacks, we sought to better understand the properties that underpin what appears to be the photonic PUF's resilience to ML-based attack and thus low susceptibility to the same.

\begin{figure} [ht]
   \begin{center}
   \begin{tabular}{c}
   \includegraphics[width=0.6\linewidth]{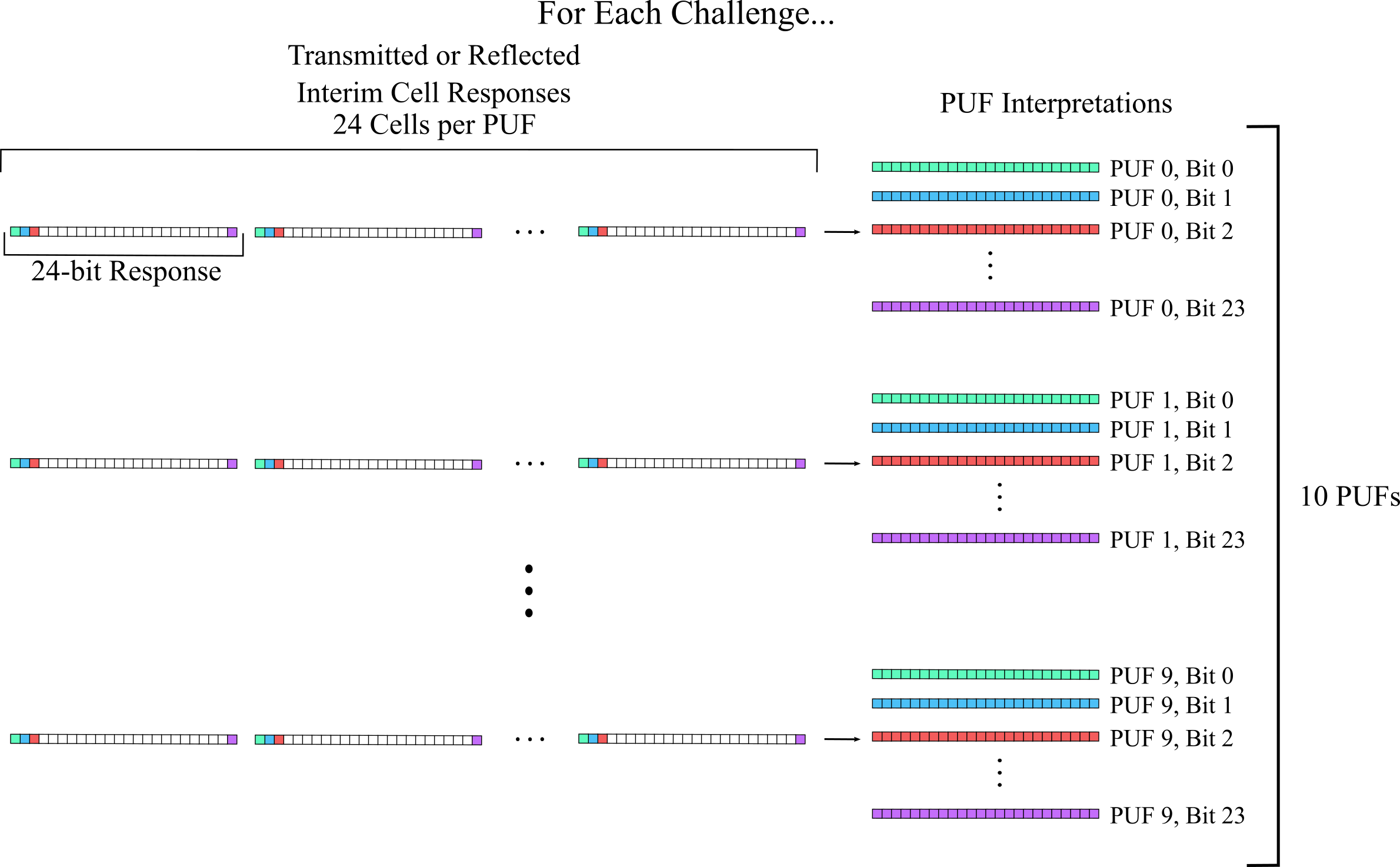}
   \end{tabular}
   \end{center}
   \caption
   { \label{fig:single_crp_data} 
Interim responses and PUF interpretation construction for a single challenge applied to ten PUFs. Interim responses from either Output $1$ or Output $2$ are used to form $24$ different PUF interpretations. The zeroth interpretation combines the bits with index zero, the first interpretation combines the bits with index one, etcetera. This process is repeated for whichever of Output $1$ or Output $2$ was not previously used. The entire process generates a total of $48$ interpretations for each of ten PUFs, providing a total of $480$ PUFs' worth of data.}
\end{figure}

\begin{figure} [ht]
   \begin{center}
   \begin{tabular}{c}
   \includegraphics[width=0.7\linewidth]{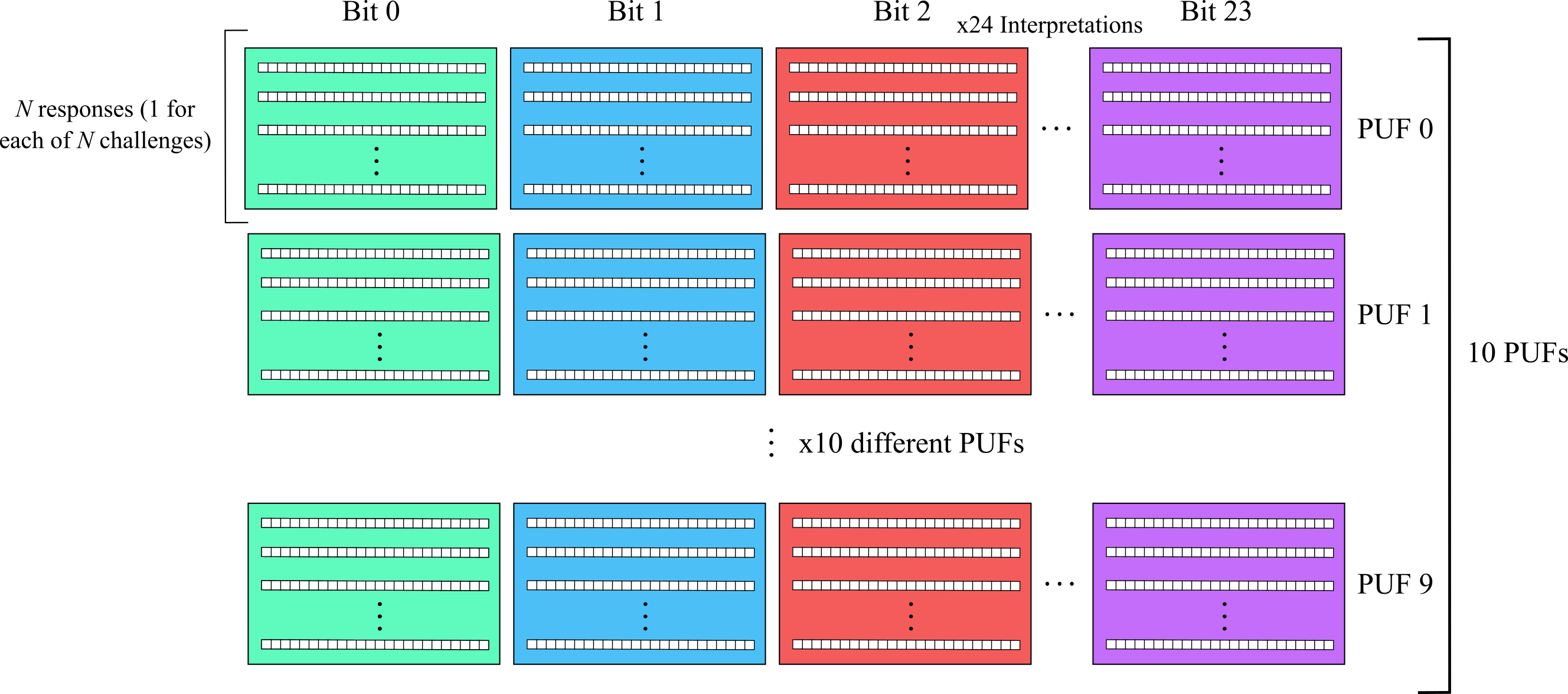}
   \end{tabular}
   \end{center}
   \caption
{ \label{fig:multiple_crp_data_puf_interpretations} 
PUF interpretations for $N$ challenges, each applied to ten different PUFs. In the datasets for this paper, $N=212$,$929$. The analysis of the following sections is applied across the ten PUFs for each PUF interpretation. For example, we analyze the ten PUFs for interpretation zero, the ten PUFs for interpretation one, etcetera. This allows for assessing which interpretations provide PUFs with the best properties.}
\end{figure}

The remainder of this section analyzes synthetic data obtained from the computational model of the photonic PUF described in Section \ref{sec:modeling_the_puf}.
We begin with a description of the data itself: we consider ten synthetic PUFs, each comprised of $24$ cells that provide interim responses.
As discussed in Section \ref{sec:modeling_the_puf}, each PUF has $48$ PUF interpretations, and we consider them all to draw conclusions about which interim response bit indices provide interpretations that are least likely to be susceptible to ML-based threats.
Figures \ref{fig:single_crp_data} and \ref{fig:multiple_crp_data_puf_interpretations} illustrate the data, with Figure \ref{fig:single_crp_data} illustrating the data for a single challenge-response pair, and Figure \ref{fig:multiple_crp_data_puf_interpretations} illustrating the data organized by PUF interpretation.

Before turning to the analysis, we make two terminological notes about the CRP bitstrings.
First, as discussed in Section \ref{sec:modeling_the_puf}, every challenge and response bitstring has two parts: one comes from the field magnitude squared ($E_x^2$), while the other comes from the phase difference ($\Delta\phi$).
This means that---for each interim response---there are \textit{two} most significant bits (MSB): bit zero is the MSB for the field magnitude squared, while bit $12$ is the MSB for the phase difference.
The same is true for the least significant bits (LSB): bit $11$ is the LSB for the field magnitude squared, and bit $23$ is the LSB for the phase difference.
So, PUF interpretations zero through five and $12$ through $17$ are constructed using more-significant bits, while PUF interpretations six through $11$ and $18$ through $23$ are constructed using less-significant bits.
We predict that PUF interpretations constructed using less-significant bits will exhibit better properties than those constructed using more-significant bits, because less-significant bits are more likely to change between responses than are more-significant bits.
At the same time, in a laboratory setting with a fabricated PIC PUF, the least significant bits will be most susceptible to measurement noise, meaning they are less likely to be reproducible or to accurately reflect the behavior of the PUF.
Consequently, there is likely a bit range that is close enough to the least significant bit to vary frequently, but far enough away to reflect more than noise.
In this work, we assess which interim bit indices lead to PUF interpretations that exhibit good properties in a synthetic data context, and ongoing work aims to pair these findings with an assessment of noise's effect when working with fabricated PIC PUFs.

For the remainder of the paper, \textit{more-significant bits} and \textit{more-significant interpretations} refer to interim response bit indices zero through five and $12$ through $17$ and to the PUF interpretations derived therefrom.
Similarly, \textit{less-significant bits} and \textit{less-significant interpretations} refer to the bit indices and interpretations six through $11$ and $18$ through $23$.

Additionally, \textit{adjacent} challenges refer to challenge bitstrings that came from Jones vectors separated by just one step.
For example, two challenges, each with $E_x^2=0.5$ but with $\Delta\phi$ values of $0.1$ and $0.187$ are adjacent, because the $\Delta\phi$ step size used to generate CRP datasets is $0.087$.
Adjacent responses will refer to response bitstrings whose corresponding challenges are adjacent.

\subsection{Challenge-Response Distribution}
\begin{figure} [ht]
   \begin{center}
   \begin{tabular}{c}
   \includegraphics[width=0.7\linewidth]{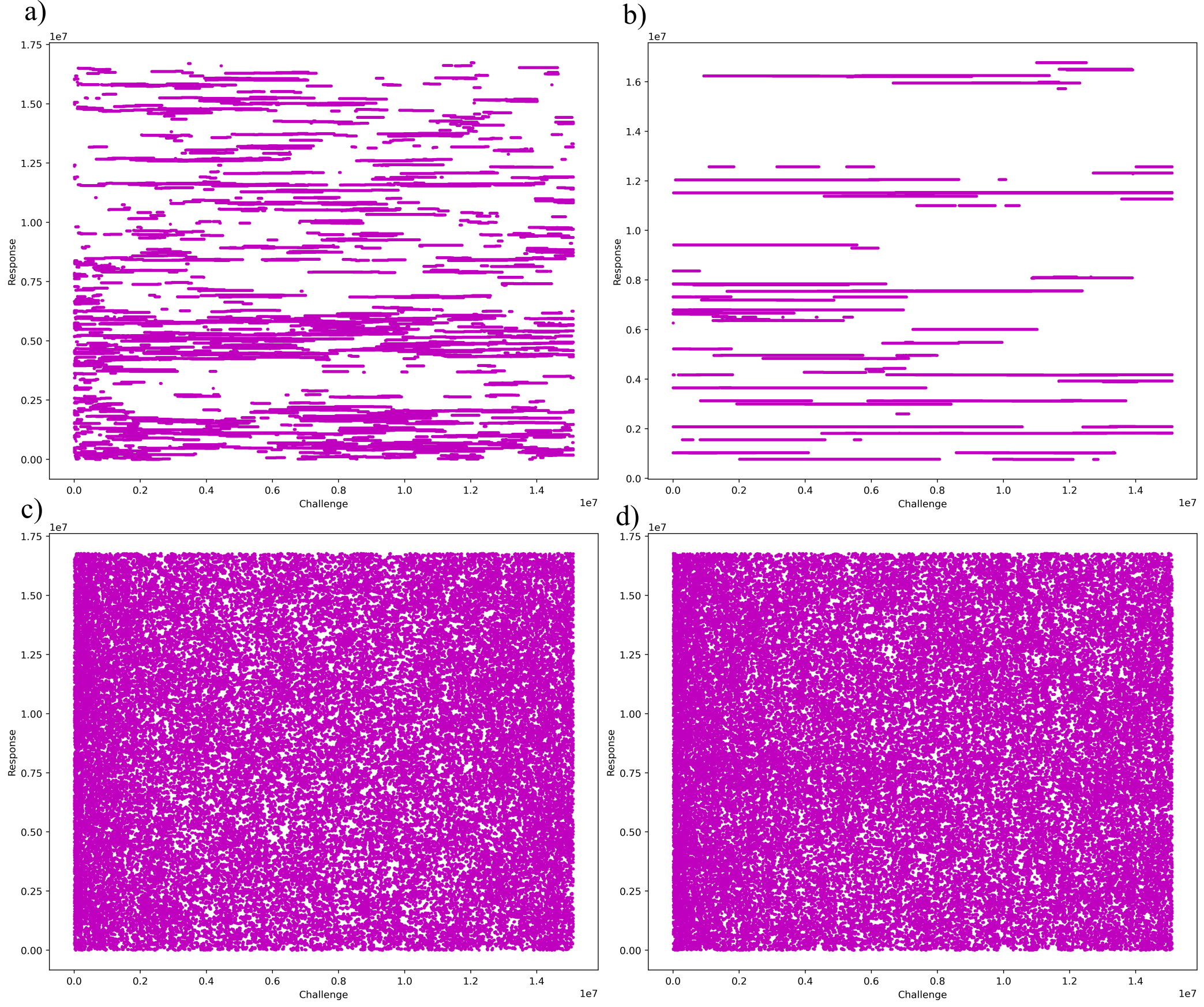}
   \end{tabular}
   \end{center}
   \caption
{ \label{fig:crp_distribution} 
Challenge-response distribution for four PUF interpretations. All are for PUF zero and Output $1$ interpretations: Subfigure a) is interpretation zero, Subfigure b) is interpretation $14$, Subfigure c) is interpretation nine, and Subfigure d) is interpretation $21$. The exhibited trend holds for all nine other PUFs considered, and for Output $2$ bit interpretations, as well. These plots are simply representative examples.}
\end{figure}

We begin with the relationship between challenges and responses.
Ideally, this relationship is not predictable, meaning that adjacent challenges should produce responses that are quite different.
This ideal translates to a scatter plot of CRP points interpreted as base-$10$ numbers, where the points are spread evenly and entirely over the range of CRPs.
Consider Figure \ref{fig:crp_distribution}, which illustrates the CRP distribution for four PUF interpretations.
In each subplot, the challenge bitstrings are interpreted as base-$10$ values and plotted on the $x$-axis, while the response bistrings are interpreted as the same and plotted on the $y$-axis.
The four plots are from PUF zero with Output $1$ interim responses, but the same trend holds for all PUFs considered.
Each Subfigure is a different PUF interpretation: Subfigure a) is interpretation four, Subfigure b) is interpretation $14$, Subfigure c) is interpretation nine, and Subfigure d) is interpretation $21$.
Thus, Subfigures a) and b) come from bits that are more-significant, while Subfigures c) and d) come from bits that are less-significant.

The trend for challenge-response distribution thus fits our prediction about which PUF intrepretations are likely to exhibit better properties.
While interpretations four and $14$ clearly have limited response variability for adjacent challenges, interpretations nine and $21$ exhibit a far more uniform distribution of CRPs, with adjacent responses spanning the entire range of $y$-axis values.

Another way to illustrate the overall relationship between responses for a given PUF interpretation is inter-response correlation.
A PUF with low inter-response correlation is ideal, because responses that are unrelated should exhibit more resilience to ML-based threats.
Indeed, low response correlation was the initial motivation for a cell-based PUF, which generally reduces the relation between CRPs.
However, that is not true for all PUF interpretations: as with the challenge-response distribution plots, we find that some interpretations are better than others.

\begin{figure} [ht]
   \begin{center}
   \begin{tabular}{c}
   \includegraphics[width=0.7\linewidth]{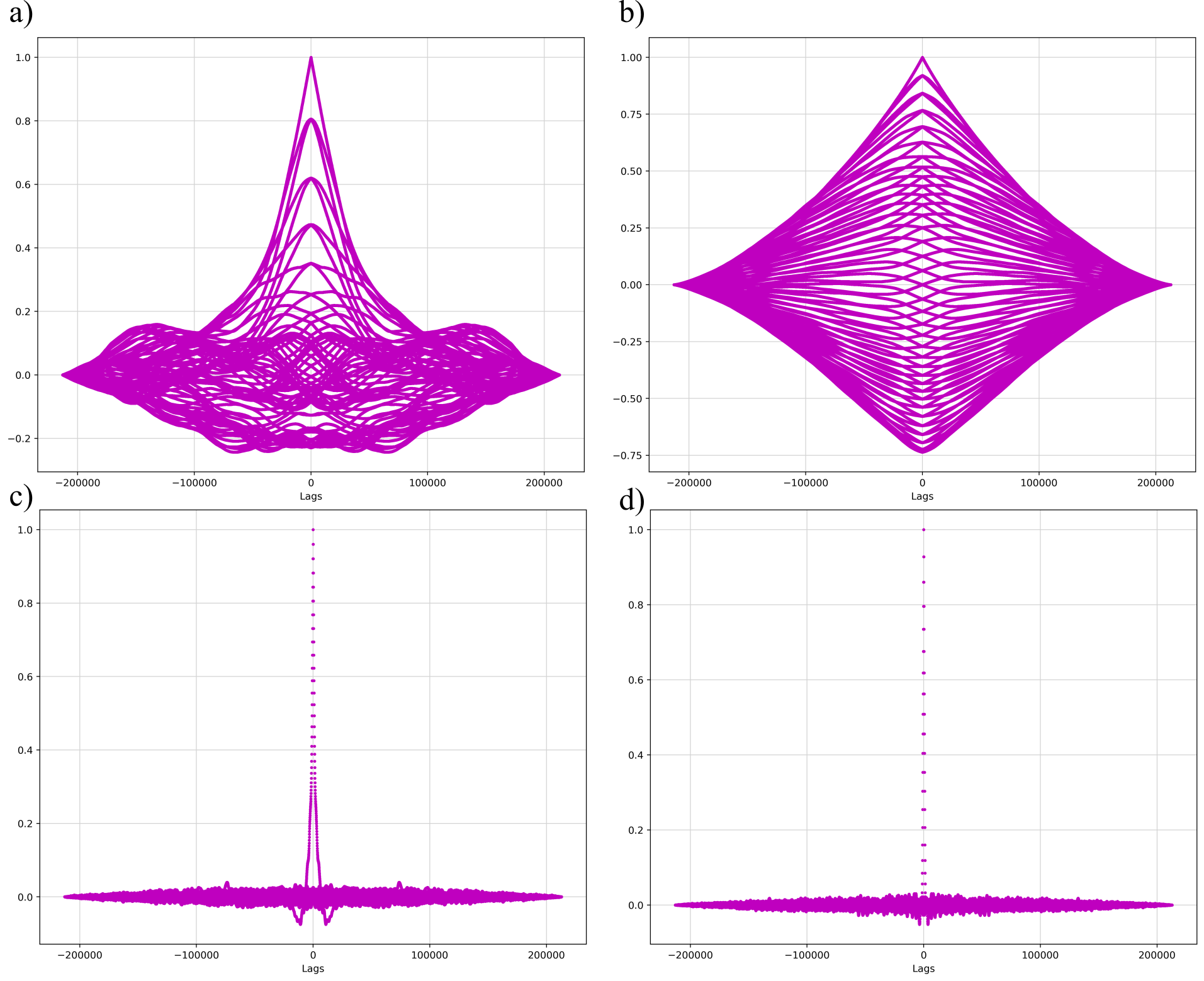}
   \end{tabular}
   \end{center}
   \caption
{ \label{fig:autocorrelation} 
Autocorrelation for four PUF interpretations. All are for PUF zero and Output $1$ interpretations: Subfigure a) is interpretation zero, Subfigure b) is interpretation $14$, Subfigure c) is interpretation nine, and Subfigure d) is interpretation $21$. The exhibited trend holds for all nine other PUFs considered, and for Output $2$ bit interpretations, as well. These plots are simply representative examples.}
\end{figure}

Figure \ref{fig:autocorrelation} consists of four autocorrelation plots that are representative of all those considered.\footnote{We use the Python \texttt{statsmodels} implementation of autocorrelation computation\cite{statsmodels}.}
Each plot shows the extent of relation between the $212$,$929$ responses for one of PUF zero's Output $1$ interpretations: Subfigure a) is for interpretation four, Subfigure b) is for interpretation $14$, Subfigure c) is for interpretation nine, and Subfigure d) is for interpretation $21$.
The extent of correlation (between zero and one, inclusive) is on the $y$-axis, and the \textit{lag}---an index for responses---is on the $x$-axis.
So, for example, if we have three CRPs (challenge $A$ and response $A$, challenge $B$ and response $B$, and challenge $C$ and response $C$), the lag between responses $A$ and $B$ is one, because responses $A$ and $B$ are adjacent, while the lag between responses $A$ and $C$ is two.
An ideal plot is a \textit{delta function}, which has a single spike with amplitude one at $x=0$ and values of zero everywhere else, indicating that any given response is correlated only with itself.  
Therefore, Subfigures a) and b) are far from ideal, while Subfigures c) and d) are much closer.
This is consistent with the above discussion of CRP distribution plots: interpretations four and $14$ are constructed with more-significant bits, so exhibit more patterning than the less-significant interpretations of nine and $21$.

\subsection{Uniqueness}
As described in Section \ref{sec:background}, uniqueness measures the amount that responses for a given challenge differ between PUFs.
We compute the uniqueness across the ten PUFs for each of the interpretations.
For example, the uniqueness across the ten PUFs with interpretation nine is $0.50004$.
This indicates that if we feed a challenge into two different PUFs, and then compare the two responses created using interim bit nine, $50\%$ of the bits will differ between those responses, on average.

Overall, most of the PUF interpretations fare well with respect to uniqueness: the average uniqueness is $0.47528$, the median uniqueness is $0.50011$, and the standard deviation is $0.10002$.
Interpretations eight through $11$, $15$ through $19$, $21$, and $22$ all exhibit `perfect' uniqueness of $0.500$ to three figures.
Except for an outlier of interpretation zero (which has a uniqueness of zero across the ten PUFs), the minimum uniqueness is $0.44211$, and the maximum uniqueness is $0.50844$.
The outlier is explainable given the zeroth bit's role as the MSB of the field magnitude squared, which is thus not likely to vary and---in the synthetic datasets considered---varies not at all.
The conclusion is thus that less-significant interpretations provide better uniqueness than their more-significant counterparts, while the most-significant interpretations can offer very poor uniqueness.
This is consistent with previous discussion suggesting that less-significant PUF interpretations exhibit preferable properties.

\subsection{Uniformity}
\begin{figure} [ht]
   \begin{center}
   \begin{tabular}{c}
   \includegraphics[width=\linewidth]{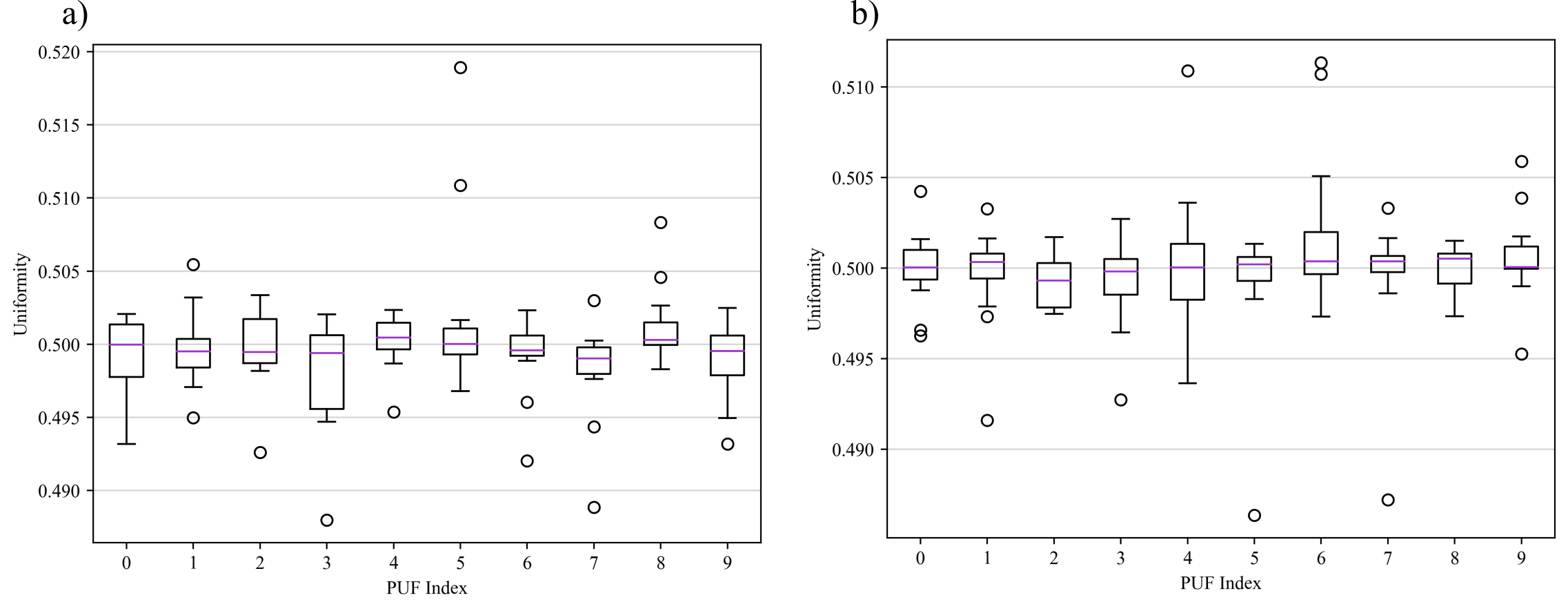}
   \end{tabular}
   \end{center}
   \caption
{ \label{fig:uniformity} 
A selection of uniformity values for ten PUFs with $24$ interpretations each. PUF index is along the $x$-axis. The uniformity value is along the $y$-axis. Each box plot contains PUF interpretations six through $11$ and $18$ through $23$. Subfigure a) is for Output $1$ responses, while Subfigure b) is for Output $2$ responses. The purple lines indicate the second quartile (median), and the dots indicate outliers that are outside $1.5\text{(IQR)}$, where $\text{IQR}$ is the interquartile range.}
\end{figure}

As described in Section \ref{sec:background}, uniformity measures how much a given PUF's responses are comprised of an equal number of zero and one bits.
We compute the uniformity for each PUF interpretation of each different PUF, for a total of $24*2*10=480$ uniformity values ($24$ interpretations for each of two categories (Output $1$ and Output $2$) for ten different PUFs).
Figure \ref{fig:uniformity} contains some of these values: Subfigure a) contains information about Output $1$ interpretations, and Subfigure b) contains the same for Output $2$ interpretations.
Each Subfigure contains ten box plots, with one for each of the PUFs.
And each box plot represents $12$ data points: the uniformity for interpretations six through $11$ and $18$ through $23$.
So, for example, the second box plot in Subfigure a) is for PUF one, and it indicates that---of Output $1$ interpretations six through $11$ and $18$ through $23$---the median uniformity is just under $0.5$, the range is $0.00625$ (not counting outliers), and there are two outliers ($0.495$ and just over $0.505$) because the interquartile range is so small that even values within $0.01$ of each other are outliers.

We present data from these PUF interpretations because the uniformities closest to the ideal of $0.5$ are consistently those from the less-significant PUF interpretations.
Specifically, the mean for Output $1$ PUFs with these interpretations is $0.49966$, and is $0.49991$ for Output $2$ PUFs.
Furthermore, the average standard deviation for the Output $1$ PUFs and these bit interpretations is $0.00230$, and for Output $2$ PUFs is $0.00038$, indicating that the uniformity is consistently very close to ideal.

Conversely, for the interpretations not plotted (zero through five and $12$ through $17$), the uniformity is worse, with an average of $0.43132$ and a standard deviation of $0.04506$ for Output $1$ interpretations, and an average of $0.42585$ and standard deviation of $0.03832$ for Output $2$ interpretations.
Additionally, some bit interpretations have consistently poor uniformities, including interpretation zero with a uniformity of zero in all cases.
This is consistent with previous discussions of the likely-more-oscillatory behavior of less-significant bits, and is thus further evidence that such interpretations provide PUFs with the best properties.

\subsection{Bit Aliasing}
\begin{figure} [ht]
   \begin{center}
   \begin{tabular}{c}
   \includegraphics[width=\linewidth]{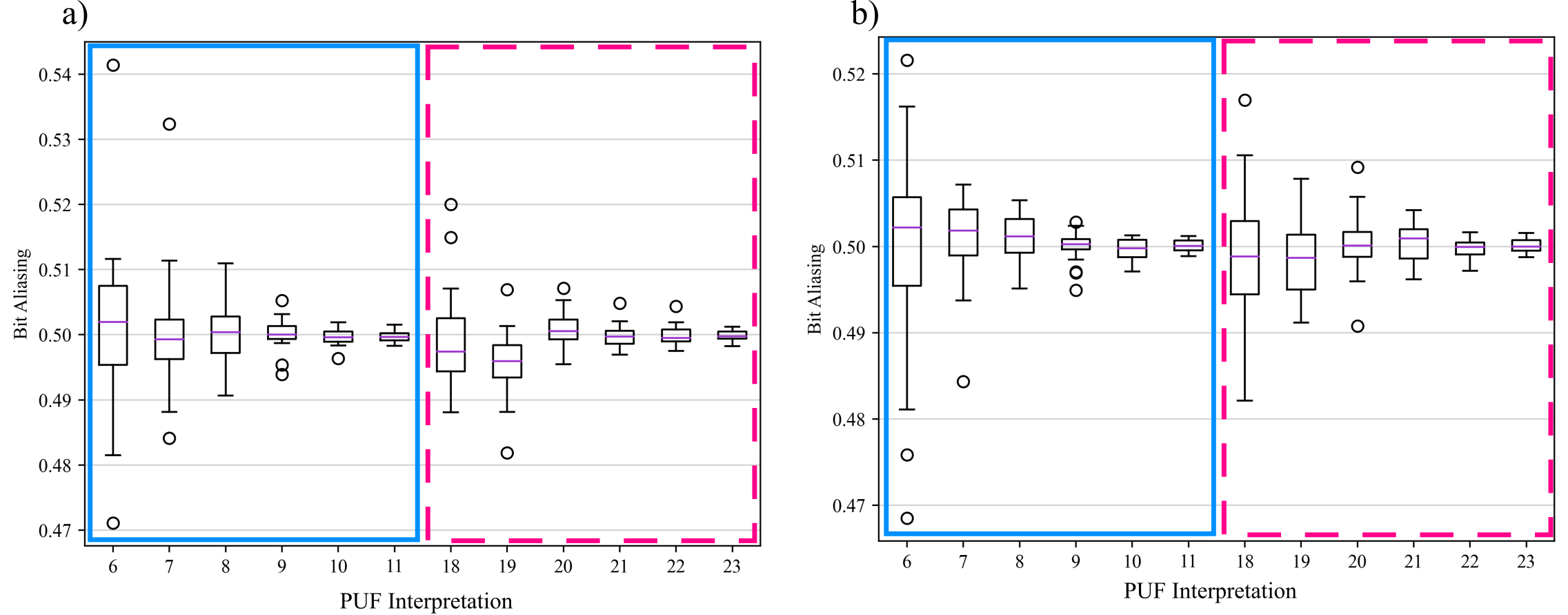}
   \end{tabular}
   \end{center}
   \caption
{ \label{fig:bit_aliasing} 
Bit aliasing for PUF interpretations six through $11$ and $18$ through $23$. Subfigure a) is for Output $1$ responses, while Subfigure b) is for Output $2$ responses. The solid blue rectangle denotes that PUF interpretations $6$ through $11$ are for field magnitude, while the dashed magenta rectangle denotes that PUF interpretations $18$ through $23$ are for phase. This explains why the range of the bit aliasing values falls from six to $11$, rises between $11$ and $18$, and then falls again from $18$ to $23$. (We move from the least-significant bit of field magnitude to the sixth-least bit of phase between $11$ and $18$.)}
\end{figure}

As described in Section \ref{sec:background}, bit aliasing measures how evenly each bit index in the responses takes on values of zero and values of one across all CRPs.
So, for each PUF interpretation, there are $24$ bit aliasing measures, showing how well the zeroth bit of the PUF interpretation is split between values of zero and one, how well the first bit of the PUF interpretation is split between values of zero and one, etcetera.
Figure \ref{fig:bit_aliasing} illustrates the results for the $24$-bits of the less-significant PUF interpretations.

Because Figure \ref{fig:bit_aliasing} has a substantial amount of information, we briefly describe one of the box plots.
Consider Subfigure a), which is from Output $1$ responses across our ten synthetic PUFs.
The solid blue rectangle surrounds data from PUF interpretations six through $11$ (\textit{i.e.}, interpretations from field magnitude bits) while the dashed magenta rectangle surrounds data from PUF interpretations $18$ through $23$ (\textit{i.e.}, interpretations from phase difference bits).
So, the first box plot (labelled as ``$6$'' on the $x$-axis) contains $24$ bit aliasing values: one for each of the bit positions in the responses for the PUF interpretations that were created by combining bit six of interim responses.  
This first box plot has a median of $0.50067$ and a standard deviation of $0.01313$.

Figure \ref{fig:bit_aliasing} is consistent with the results discussed to this point: the average bit aliasing and standard deviation changes from $0.50067$ and $0.01313$ with Output $1$ PUF interpretation six to an average of $0.49976$ and a standard deviation of $0.00087$ for PUF interpretation $11$.
Therefore, we reach the same conclusion as above: the less-significant PUF interpretations provide better PUF properties.

\section{Conclusion and Future Work}
We demonstrate that synthetic data from the computational model of the photonic PUF exhibits CRP-distribution, uniqueness, uniformity, and bit aliasing of a robust PUF.
Specifically, we show that the PUF interpretations comprised of less-significant bits (interpretations six through $11$ and $18$ through $23$) best fit the requirements of ideal uniqueness, uniformity, and bit aliasing, and also exhibit a more uniform CRP distribution and less correlated responses than more-significant PUF interpretations.
This work thus supports the findings about ML-threat resilience of the synthetic photonic PUF datasets.

This paper also raises several avenues for further work, many of which are in progress.
On the model side, we have not addressed how many cells are required to achieve uncorrelated responses and associated ideal PUF properties.
This work considered the `limiting' case of one interim bit per cell, but it is possible that we might select a few bits from each cell while avoiding correlation or damage to the uniqueness, uniformity, or bit aliasing.
We might thus be able to create an equally strong PUF with fewer cells, which would simplify PIC fabrication.
As mentioned above, we are also assessing the efficacy of additional ML-based attacks, and we anticipate that the photonic PUF will illustrate resilience to these as well.
Finally, we have begun experiments to connect analysis of synthetic data from the computational model with CRP datasets from fabricated photonic PUFs.
This will allow for property analysis that is not practical for the computational model (\textit{e.g.}, for assessing reliability), while also allowing for exploration of the effect of noise on choosing the interim response bits with which to create PUF interpretations.
(For example, the least-significant bits---$11$ and $23$---may be too beset by instrument-related noise to be useful for forming PUF interpretations, even as the computational model showed that the least-significant PUF interpretations exhibit the best PUF properties.

\appendix

\bibliography{references} 
\bibliographystyle{spiebib}

\end{document}